\documentclass[12pt]{article}
\usepackage{amsfonts,epsfig,latexsym,amscd}
\usepackage{amssymb}

\newcommand{\R}{{\mathbb{R}}}
\newcommand{\Z}{{\mathbb{Z}}}

\newcommand{\C}{{\mathbb{C}}}
\newcommand{\CP}{{\mathbb{C}}{{P}}}
\newcommand{\I}{{\mathbb{I}}}

\newcommand{\be}{\begin{equation}}
\newcommand{\ee}{\end{equation}}
\newcommand{\bea}{\begin{eqnarray}}
\newcommand{\eea}{\end{eqnarray}}
\newcommand{\bean}{\begin{eqnarray*}}

\newcommand{\eean}{\end{eqnarray*}}
\font\upright=cmu10 scaled\magstep1

\newcommand{\PP}{\hbox{\upright\rlap{I}\kern 1.5pt P}}

\newcommand{\identity}{{\upright\rlap{1}\kern 2.0pt 1}}

\newcommand{\HH}{\mbox{\hbox{\upright\rlap{I}\kern 1.7pt H}}}

\newcommand{\fr}{\frac}

\newcommand{\sg}{\sigma}

\newcommand{\pr}{\partial}
\newcommand{\x}{ {\bf x} }
\newcommand{\hs}{\hspace{5mm}}

\newcommand{\dg}{\dagger}

\newcommand{\ve}{\varepsilon}
\newcommand{\acc}{\\[3mm]}

\newcommand{\vv}{{\bf v}}

\begin{document}
\setcounter{page}{0}
\begin{titlepage}
\strut\hfill
\vspace{0mm}
\begin{center}

{\large\bf  BPS Monopoles \& Open Spin Chains }
\vspace{12mm}

{\bf Anastasia Doikou${}^*$ \ and \ Theodora Ioannidou${}^\dg$}
\\[8mm]
\noindent ${}^*${\footnotesize Department of Engineering Sciences, University of Patras,
GR-26500 Patras, Greece }\\
{\footnotesize {\tt E-mail: adoikou@upatras.gr}}
\\[8mm]
\noindent ${}^\dg${\footnotesize Department of Mathematics, Physics and Computational Sciences, Faculty of Engineering,\\
Aristotle University of Thessaloniki, GR-54124 Thessaloniki, Greece }\\
{\footnotesize {\tt E-mail: ti3@auth.gr}}

\vspace{12mm}

\begin{abstract}
We construct  $SU(n+1)$ BPS spherically symmetric monopoles with minimal symmetry breaking by solving the full Weyl  equation.
In this context,  we explore  and discuss  the existence  of open spin chain-like part within the Weyl equation.
For instance, in the $SU(3)$ case the relevant spin chain is the $2$-site spin $1/2$ $XXX$ chain with open boundary conditions. We exploit the existence of such a spin chain part in order to solve the full Weyl equation.

\noindent

\end{abstract}

\end{center}
\end{titlepage}


\section{Introduction}

The integrability of the self-dual Yang-Mills equation was first realized by Ward \cite{W}, who demonstrated that the twistor transform of Penrose could be used to provide a correspondence between instantons and certain holomorphic vector bundles over the twistor space  $CP^3$ \cite{WW}.
Since then many alternative methods (see \cite{MS} and References therein) have been proposed in order to find instanton and monopole solutions of the self-dual Yang-Mills equation.

A powerful approach introduced by Nahm \cite{Nahm,Nahm2} is the so-called Atiyah-Drinfeld-Hitchin-Manin-Nahm (ADHMN) construction.
The Atiyah-Drinfeld-Hitchin-Manin (ADHM) approach \cite{ADHM}, allows the construction of instantons in terms of linear algebras in a vector space, which dimension is related to the instanton number.
Since monopoles correspond to infinite action instantons, an adaption of the ADHM construction involving an infinite dimensional vector space might also be  possible.
Nahm was able to formulate such an adaption in the ADHMN construction.
To perform this construction, a nonlinear ordinary differential equation (i.e. the Nahm equation) must be solved and its solutions (i.e. the Nahm data) are used to define the Weyl equation.

The Nahm equations provide a system of non-linear ordinary differential equations for three $n\times n$ anti-hermitian matrices $T_i$ (the so-called Nahm data) functions
of the variable $s$:
\be
\fr{dT_i}{ds}=\fr{1}{2}\, \ve_{ijk}\,[T_j,\ T_k]
\label{Nahm}
\ee
 where $n$ is the magnetic charge of the BPS monopole configuration and
 $\ve_{ijk}$ is the totally antisymmetric tensor.

In the ADHMN approach, the construction of $SU(n+1)$ monopole solutions of the Bogomolny equation with topological charge $n$ is translated to the following problem,  known as the {\it inverse Nahm transform} \cite{Nahm}. Given the Nahm data for a $n$-monopole the one-dimensional Weyl equation
\be
\left( \I_{2n}\,\fr{d}{ds}-\I_n\otimes x_j \sg_j +iT_j\otimes\sg_j\right)\vv({\bf x},s)=0
\label{Weyl}
\ee
for the complex $2n$-vector $\vv(\x,s)$, must be solved. $\I_n$ denotes the $n\times n$ identity matrix, ${\bf x}=(x_1,x_2,x_3)$ is the position in space at which the monopole fields are to be calculated and  ($\sigma_1,\sigma_2,\sigma_3)$  are the familiar $2 \times 2$ Pauli matrices.
In the minimal symmetry breaking, the Nahm data $T_i'$s can be cast as
\be
T_i =-{i\over 2}\,f_i\,\tau_i, \hs i=1, 2,3, \label{data}
\ee
where $\tau_i$'s form  the $n$-dimensional representation of $su(2)$ and satisfy
\be
[\tau_i,\ \tau_j] = 2i \varepsilon_{ijk}\tau_k.
\ee

Let us choose an orthonormal basis for these solutions, satisfying
\be
\int \hat \vv^\dg \hat \vv \,ds=\I. \label{wnor}\ee
Given $\hat \vv(\x,s)$, the normalized vector computed from (\ref{Weyl}) satisfying (\ref{wnor}), the Higgs field $\Phi$ and gauge potential $A_k$ are given by
\bea
 &&\Phi=-i\int s\, \hat \vv^\dg \hat \vv \,ds \nonumber\\
&& A_k=\int \hat \vv^\dg \,\pr_k \hat \vv\, ds,\label{Higgs}
\eea
where the integrations are to be performed over the range spanned by the minimum and maximum
eigenvalues of the asymptotic form of $\Phi$.
Then the corresponding  Higgs field and gauge potentials satisfy the self-dual Yang-Mills equations, and they are smooth functions of $\x$.

Recently in \cite{doikou-ioannidou}, $SU(n+1)$ (for generic values of $n$) {\it spherically symmetric} BPS monopoles with minimal symmetry breaking located at the origin were obtained by solving the  Weyl equation for ${\bf x}=(0,0,r)$ (see also \cite{WGA} for generic spherical monopoles).
In this case, the corresponding fields (\ref{Higgs}) were calculated along the axis $(0,0,r)$.
However, their azimuthal dependence can not be implemented in a straightforward manner.
To do so either the full Weyl equation for ${\bf x}=(x,y,z)$  should be solved or, a suitable similarity transformation should be introduced in order to derive the corresponding solutions in terms of the simple ones of \cite{doikou-ioannidou}.
That way {\it all}  the gauge fields can be recovered.

As will be transparent in the next section, solving the full Weyl equation for the generic case is a rather technically difficult task.
However, the crucial observation is the existence of a spin chain-like part
within the full Weyl equation.
This observation allows us to bypass the existing technical difficulties.
More precisely, we present a new approach for constructing the spherically symmetric monopole solutions of the self-dual Yang-Mills equation by connecting the corresponding full Weyl equation with a one-dimensional quantum integrable model, that is  the $XXX$ spin chain (also knows as, {\it isotropic Heisenberg model}).
To our knowledge this is the first attempt to make such a connection; although, the integrability of the Weyl equation was also discussed in \cite{TTX}.

In fact, it is worth noting that the effective Yang-Mills dynamics in several special limits is described by completely integrable systems  related to Heisenberg spin chain and its generalization (see, e.g. \cite{BBGK} and References within).
It was shown for instance in \cite{MZ}, that the one loop mixing matrix  for anomalous dimensions in $N=4$ super-symmetric Yang-Mills theory can be identified with the conventional isotropic Heisenberg spin $1/2$ chain.
Also, it was shown in \cite{Lipatov, FK} that the $XXX$ Heisenberg chain of noncompact spin $s=0$,
describes the high energy scattering of hadrons in multi-color Quantum Chromo-Dynamics \cite{Lipatov}.

\section{BPS Monopoles}
\subsection{$SU(2)$ BPS Monopoles}

Recall the full Weyl equation (\ref{Weyl}) takes the form
\be
\left ({d \over ds } + i T_j \otimes \sigma_j - x_j \otimes \sigma_j \right ) \vv(\x,s) =0, \label{su2}
\ee
where  $(x_1,x_2,x_3)=(x,y,z)$ and  the vector $\vv(\x,s)$ is of the form $\vv=  \left( \begin{array}{c}
v_1\\v_2\end{array} \right)$.

The spin zero representation of the $su(2)$ case is trivial (that is, $T_i=0$).
Therefore, the corresponding full Weyl equation (\ref{su2}) for the $SU(2)$ BPS monopole reduces to:
\be
{d \over ds} \left( \begin{array}{c}
	              v_1\\
				   v_2
\end{array} \right) - \left( \begin{array}{cc}
	               r u & \  \  \ r \psi\\
				   r \bar \psi & -r u
\end{array} \right) \left( \begin{array}{c}
	              v_1\\
				   v_2
\end{array} \right) =0,\label{equ0}
\ee
where we define $u = \cos\theta$, $\psi = \sin\theta\, e^{-i \phi}$ and its complex conjugate $\bar{\psi}$ and thus,
\be
z=ru,\ ~~~~ x - iy=r \psi,\ ~~~~ x+iy=r \bar \psi.
\label{polar}
\ee
Notice that $u^2+|\psi|^2=1$.
Then, equation (\ref{equ0}) reduces to the following set of differential equations
\bea
&& \dot{v}_1 - ru\, v_1 - r \psi \,v_2 =0 \nonumber\\
&& \dot{v}_2 + r u \,v_2 - r \bar \psi \,v_1 =0, \label{v1}
\eea
and its solution  is given by
\bea
&& v_1 = \kappa_1 \,e^{rs} + \kappa_2 \,e^{-rs} \nonumber\\
&& v_2 = {\kappa_1 \over \psi}\left(1-u\right) e^{rs} -{\kappa_2 \over \psi} \left(1+u\right) e^{-rs}.
\label{s2}
\eea
Next choose a  vector $\tilde \vv$ to be of the form (\ref{s2}) with constants $\tilde \kappa_1,\ \tilde \kappa_2$, such that the set $\{\vv,\ \tilde \vv\}$ defines an orthonormal system.
Thus, introducing the scalar product
\be
<\vv, \tilde \vv >= \int_{-1}^1  \vv^{\dag}(s) \tilde \vv(s)\,ds,
\ee
 the following relations must be satisfied
\be
<\vv, \vv> \,= \,<\tilde \vv,\tilde \vv> \,= 1, ~~~~ < \vv, \tilde  \vv> \,=0.\label{nor}
\ee
Then,  due to  the constraints  (\ref{nor}),  the solutions (\ref{s2})  lead to the following choice of constants
\bea
&&\kappa_1 = \sqrt{r \over 4 \sinh 2r}\ \left(1+u\right), \hs~~~~~\kappa_2 =\sqrt{r \over 4 \sinh 2r}\ \left(1-u\right), \nonumber\\
&& \tilde \kappa_1 = \sqrt{r \over 4 \sinh 2r}\ \psi,\hs \hs\ \hs\,\,\hs \tilde \kappa_2 =-\sqrt{r \over 4 \sinh 2r}\ \psi.
\eea
Therefore, we end up with  two orthonormal vectors of the form
\bea
&&\vv =\sqrt{r \over 4 \sinh 2r}\left( \begin{array}{c}
	           \left(1+u\right)e^{rs}+\left(1-u\right)e^{-rs}  \\
				   \bar \psi\, e^{rs} - \bar \psi\, e ^{-rs}
\end{array} \right),  \nonumber  \acc
&&\tilde \vv= \sqrt{r \over 4 \sinh 2r}\left( \begin{array}{c}
	              \psi\, e^{rs} - \psi \,e^{-rs}\\
\left(1-u\right)e^{rs} +\left(1+u\right)e^{-rs}
\end{array} \right). \label{gen1}
\eea
Any other choice of constants is gauge equivalent  to the above ones (due to gauge freedom).

Let $\vv = \vv_1$ and  $\tilde \vv =  \vv_2$.
Then, the associated Higgs field and gauge potentials are  $2 \times 2$ matrices with elements
\bea
&&\Phi_{ij} = -i \int_{-1}^1 s \,\vv_i^{\dag} \vv_j\ ds  \nonumber\\
&& A_{k_{ij}}=\int_{-1}^1  \vv_i^{\dag}\, \pr_k \vv_j\ ds. \label{hig1}
\eea

\subsection{$SU(n+1)$ BPS Monopoles}

The $n$-dimensional representation of $su(2)$ is of the form
\be
\tau_1 = \sum_{k=1}^{n-1} C_k \left(e^{(n)}_{k k+1}+ e^{(n)}_{k+1 k}\right),
\hs \tau_2=i\sum_{k=1}^{n-1} C_k \left( e^{(n)}_{k+1 k}- e^{(n)}_{k k+1}\right),\hs
\tau_3 = \sum_{k=1}^na_k\, e^{(n)}_{kk}
\ee
where $e^{(n)}_{ij}$ are $n\times n$ matrices defined by: $\left(e^{(n)}_{ij}\right)_{kl} = \delta_{ik} \,\delta_{jl}$ and
\be
a_k = n+1 -2k, \hs \hs C_k =\sqrt{k\,(n-k)}.
\ee

The Nahm data for the $SU(n+1)$ spherically symmetric monopoles of charge $n$ are given by (\ref{data}) where
$~f_i = f= -{1\over s}$.
Assume that the vector $\vv(\x,s)$ is of the form
\be
\vv(\x,s) = \sum_{l=1}^n h_l(r,s)\ \hat e^{(n)}_{l} \otimes \left(g_1(r,s)
\ \hat e_{1}^{(2)} + g_2(r,s)\ \hat e_{2}^{(2)}\right)
\ee
where $\hat e_{k}^{(n)}$ is the $n$-dimensional column vector with one at the position $k\in \Z^+$ and zero elsewhere, i.e. the standard basis of $\R^n$.

Then, the Weyl equation (\ref{Weyl}) takes the form
{\small \bea
 \left [{d\over ds} +{f\over 2} \sum_{k=1}^{n-1}\, C_k\left(e_{k k+1}^{(n)} + e^{(n)}_{k+1 k}\right) \otimes \left(e_{12}^{(2)}
+ e_{21}^{(2)}\right) -
{f\over 2} \sum_{k=1}^{n-1}C_k \left(e^{(n)}_{k+1 k}-e^{(n)}_{kk+1}\right)\otimes\left(e_{21}^{(2)} -e_{12}^{(2)}\right)\right.
 \nonumber\\
\left. +{f\over 2}\sum_{k=1}^n a_k\, e_{kk}^{(n)} \otimes \left(e_{11}^{(2)} -e_{22}^{(2)}\right) \!- z \,\I \otimes
\left(e_{11}^{(2)} - e_{22}^{(2)}\right)\!- x \, \I \otimes \left(e_{12}^{(2)} +e_{21}^{(2)} \right ) \!- i y\,
 \I \otimes \left (e_{21}^{(2)} -e_{12}^{(2)} \right)\!
\right ] \nonumber\\
\sum_{l=1}^n h_l\, \hat e_{l}^{(n)} \otimes \left(g_1\, \hat e_{1}^{(2)} + g_2\,\hat e_{2}^{(2)}\right) =0.  \label{w1}
\eea}
To proceed with our computation we exploit the following properties
\be
e^{(n)}_{ij}\ e^{(n)}_{kl} = \delta_{kj}\, e_{il}^{(n)},\hs \hs e_{ij}^{(n)}\ \hat e_k^{(n)} = \delta_{jk}\, \hat e^{(n)}_i.
\ee
With the use of the latter identities, and after setting
\be
v_l(r,s) = h_l(r,s)\,g_1(r,s), \hs w_l(r,s) = h_l(r,s)\,g_2(r,s),
\ee
equation (\ref{w1}) is equivalent to the following first-order system of differential equations
\bea
&&\dot{v_1} -\left({1\over 2s}\,a_1+z\right)v_1 - \left(x-iy\right) w_1 =0, \nonumber\\
&& \dot{v}_{k+1} -{1\over s} C_{k}\,w_{k} -\left({1\over 2s}\, a_{k+1} +z\right) v_{k+1} -\left(x-iy\right)w_{k+1} =0, \nonumber\\
&& \dot{w}_{k} -{1\over s}\, C_{k} \,v_{k+1} + \left({1\over 2s} \,a_{k} +z\right) w_{k} - \left(x+iy\right)v_k=0,\nonumber\\
&& \dot{w}_n + \left({1\over 2s}\, a_n +z\right)w_n - \left(x+iy\right)v_n=0, \label{equations}
\eea
where  $k=1,2, \dots, n-1$.
Here, $\dot{v_i}$ and $\dot{w_i}$ for $i=1,\dots,n$
are the total derivatives of the functions $v_i(r,s)$ and $w_i(r,s)$
with respect to the argument $s$.

Solving the aforementioned system  is the first step in reconstructing the solution of the self-dual Yang-Mills equation from Nahm data.  Then the problem of recovering the Higgs field and gauge potentials given by (\ref{Higgs}) is linear.
However, solving the system (\ref{equations}) is a rather technically difficult task.
In the next section, we show how the existing technical difficulties can be overcome by connecting the corresponding full Weyl equation (\ref{Weyl}) with the isotropic Heisenberg model.

\section{Weyl Equation \& Open Spin Chains}

The Weyl equation (\ref{Weyl}) can be identified as a Hamiltonian system.
In particular, one can observe that (\ref{Weyl}) can be described by
 a Hamiltonian containing some bulk {\it spin-spin} interaction  and a {\it boundary term}.

For simplicity, let us  focus on the $SU(3)$ case.
The corresponding results can be extended in all other cases; however, not in a straightforward manner.
After implementing the simple Nahm data of the minimal symmetry breaking (\ref{data}), the corresponding full Weyl equation (\ref{Weyl}) can be written in the form
\be
 \left( {d \over ds} -{\cal H} \right) \vv =0 \label{weyl1}\ee
where
\be
{\cal H} = {1\over 2s} \left (\sigma_x \otimes \sigma_x + \sigma_y \otimes \sigma_y + \sigma_z \otimes \sigma_z \right ) +
x\, {\mathbb I} \otimes \sigma_x + y\, {\mathbb I} \otimes \sigma_y + z\, {\mathbb I}\otimes\sigma_z \label{h}
\ee
which is nothing else but the open $XXX$ (Heisenberg) spin chain.
The first term in (\ref{h}) corresponds to the spin-spin interaction; while the rest descibes the boundary interaction.
This is a quantum integrable model and the corresponding Hamiltonian is immediately
obtained from the so-called {\it open transfer matrix}.
We shall not provide the details of this construction in the present investigation, but we refer the interested reader to the original work discussed in \cite{sklyanin}.

The boundary part of the $XXX$ Hamiltonian (\ref{h}) can be explicitly expressed as
\be
 x \sigma_x + y \sigma_y + z \sigma_z=  z \left( \begin{array}{cc}
	               1  & \tan \theta\, e^{- i \phi} \\
				   \tan \theta\, e^{ i \phi}  & -1
\end{array} \right). \label{boundary}
\ee
It is clear that there is an one to one correspondence between the coordinates and the boundary parameters of the open $XXX$ model.

The key point in solving the generic problem  described by  (\ref{weyl1}) and (\ref{h}) is the following observation:
A similarity transformation $U$ (see, e.g. \cite{annecy-group}) exist,  which diagonalizes the boundary contribution in (\ref{h}), but leaves the bulk spin-spin  interaction invariant.
More precisely, let
\be
U= \left( \begin{array}{cc}
	               1 + \cos \theta   & \hs \sin\theta\, e^{- i \phi} \\
				    \left(1 - \cos \theta \right)e^{ i \phi}  & -\sin\theta
\end{array} \right).
\label{U}
\ee
Then it is straightforward to show that, the boundary term is transformed as
\be
U\left( \begin{array}{cc}
	               1  & \tan\theta\ e^{- i \phi} \\
				   \tan \theta\,\ e^{ i \phi}  & -1
\end{array} \right)U^{-1}= {1\over \cos\theta} \left( \begin{array}{cc}
	               1  & 0\\
				   0  & -1
\end{array} \right)
\ee
while the bulk spin-spin interaction term remains unaffected:
\be
U \otimes U \left( \sigma_x \otimes \sigma_x + \sigma_y \otimes \sigma_y + \sigma_z \otimes \sigma_z   \right ) U^{-1} \otimes U^{-1} =
\sigma_x \otimes \sigma_x + \sigma_y \otimes \sigma_y + \sigma_z \otimes \sigma_z.\label{tran}
\ee

When the transformation (\ref{U}) acts to the full Weyl equation (\ref{weyl1}) and (\ref{h}),  one gets:
\be
U \otimes U\left ( {d \over ds} - {\cal H}\right) \vv =0 \ \Rightarrow \left( {d\over ds}- {1\over 2s} \,\sigma_i \otimes\sigma_i -
r \, {\mathbb I} \otimes \sigma_z \right) \left(U \otimes U\right) \vv=0
\ee
which is nothing else but the Weyl equation with only   {\it diagonal boundary} terms.
In \cite{doikou-ioannidou}, solutions of the aforementioned equation have been obtained.
Therefore,  the inverse similarity transformation can be performed in order to  solve explicitly
the full {\it non-diagonal} problem.

Let $\vv_0$ to be the solution of the {\it diagonal problem} \cite{doikou-ioannidou}.
Then the solutions $\vv$ of the full problem can be obtained from $\vv_0$ due to the relation
\be
\vv = \left(U^{-1} \otimes U^{-1}\right) \vv_0.\label{fin}
\ee

Let us also briefly consider the simplest case, i.e. the $SU(2)$ BPS monopole.
In this case, the corresponding Hamiltonian  is just the boundary term (\ref{boundary}).
This Hamiltonian emerges from the open transfer matrix as an $1$-site open $XXX$ chain. After implementing the similarity transformation (\ref{U}) one gets:
\be
U \left({d \over ds} -{\cal H}\right) \vv =0 \ \Rightarrow \left({d\over ds} - r\, \sigma_z\right)U\, \vv =0.
\ee
Therefore,  the vector $\vv$ in terms of the solution of the diagonal problem is of the form
\be
\vv = U^{-1}\,\vv_0\label{su2v0}.
\ee

In general, for the $SU(n+1)$ case the corresponding Hamiltonian has the form:
\be
{\cal H} = {1\over 2s} \,\tau_i \otimes \sigma_i + x_i \,{\mathbb I} \otimes \sigma_i.
\ee
Assume that there exist a transformation ${\cal U}$ (ie. $n\times n$ matrix) such that the {\it bulk} interaction term is
\be
{\cal U}\otimes U\left(\tau_i \otimes \sigma_i\right)\, {\cal U}^{-1} \otimes U^{-1} = \left(\tau_i \otimes \sigma_i\right).\label{tran1}
\ee
Then, the aforementioned procedure can be applied in order to obtain all the solutions of the general $SU(n+1)$ case.
The generic solution of the Weyl equation will then be
\be
\vv =\left(\, {\cal U}^{-1} \otimes U^{-1}\right) \vv_0,
\ee
where $\vv_0$ the solution of the {\it diagonal} case found in \cite{doikou-ioannidou}.
Hence, the main question to be investigated  is the derivation of the transformation ${\cal \ U}$.
Extending our work on the $SU(2)$ and $SU(3)$ it is obvious to see that the $n\times n$ matrix ${\cal \ U}$ in the general case $SU(n+1)$ is of the form
\be
{\cal \ U}^{-1}=\left(f,(1+|\xi|^2)^{\fr{n-1}{2}}\fr{P_+ f}{|P_+f|}, \cdots, (1+|\xi|^2)^{\fr{n-1}{2}}\fr{P^{n-1}_+ f}{|P_+^{n-1} f|}\right)\label{UU}
\ee
where $f$  is an $n$-component column vector which is a holomoprhic function of $\xi=\tan(\theta/2)e^{i\phi}$ of  degree $(n-1)$, the so-called {\it harmonic maps from $S\sp2$ to $CP\sp{n}$}.
 These  spherically symmetric maps into $\CP^{n}$  are given by
\be
f=\left(f_0,...,f_j,...,f_{n-1}\right)^t, \ \ \mbox{where} \ \ f_j=\xi^j\sqrt{{n-1}\choose j}
\label{smap}
\ee
and $n-1\choose j$ denote the binomial coefficients.
It can be shown that these maps, which describe the instanton and non-instanton solutions of the $\CP^n$ model  \cite{Za}, are spherically symmetric in the sense that a rotation in $\R^3$ exists which is realized as M\"obius transformation.

For $n=2$ these are all the finite action solutions, but for $n>2$ there are other non-instanton
solutions. These can be described by introducing the operator $P_+$ defined  by its action
on any vector $f\in \C^{n}$ as
\be
P_+ f=\pr_\xi f- \fr{f \,(f^\dg \,\pr_\xi f)}{|f|^2}
\ee
and then define further vectors $P_+^k f$ by induction:
$P_+^k f=P_+(P_+^{k-1} f)$.

To proceed further we note the following useful properties of
$P_+^k f$ when $f$ is holomorphic:
\begin{eqnarray}
\label{bbb}
&&(P_+^k f)^\dg \,P_+^l f=0, \hs \hs \hs k\neq l\nonumber\acc
&&\pr_{\bar{\xi}}\left(P_+^k f\right)=-P_+^{k-1} f \fr{|P_+^k
f|^2}{|P_+^{k-1} f|^2},
\hs \hs
\pr_{\xi}\left(\fr{P_+^{k-1} f}{|P_+^{k-1} f|^2}\right)=\fr{P_+^k
f}{|P_+^{k-1}f|^2}.
\label{aaa}
\end{eqnarray}
These properties either follow directly from the definition of $P_+$
or are easy to prove \cite{Za}.
Applying $P_+$ a total of $n-1$ times to a holomorphic vector gives an anti-holomorphic
vector, so that a further application of $P_+$ gives the zero vector.
The modulus of the corresponding vector $P_+^k f$ for $f$ of the above form is
\be
\vert P\sp{k}_+f\vert\sp2=\alpha(1+\vert \xi\vert\sp2)\sp{n-2k-1},\hs k>1
\ee
where $\alpha$ depends on $n$ and $k$ \cite{IPZ}; while from construction $|f|^2=(1+|\xi|^2)^{n-1}$.
 {\bf Remark:} In order our matrix ${\cal U}$ to leave  the bulk interaction unaffected, ie. to satisfy the equation (\ref{tran1}), the norm of each  harmonic map should be equal to $(1+|\xi|^2)^{n-1}$. Thus we have to normalize analogously and obtain (\ref{UU}). (See also,  Appendix, for some specific examples).

Finally, one needs to implement the orthonormality conditions to the solutions $\vv({\bf x},s)$.
The idea is to pick $n+1$ distinct solutions and require that they satisfy the orthonormality relations:
\be
<\vv_i,  \vv_j> = \delta_{ij}. \label{ortho}
\ee
This way one may eventually identify the Higgs field and gauge potentials for the generic situation via (\ref{Higgs}).
In the next section two specific examples, displaying explicitly
how the proposed methodology works, are discussed in detail.

\section{Explicit Examples}

\subsection{ $SU(2)$ Case}
Let us first consider the simplest case within the frame we described above, i.e. the $SU(2)$ gauge.
Recall that,  the solution of the {\it diagonal} Weyl equation is of the form
\bea
\vv_0 &=& \left( \begin{array}{c}
      g_1(r,s)\\
      g_2(r,s)
\end{array} \right)\nonumber\acc
&=& \left( \begin{array}{c}
      a(r)\,e^{rs}\\
      b(r)\,e^{-rs}
\end{array} \right) .\label{diag}
\eea
Then, the corresponding solution of the full Weyl equation, that is after performing the inverse similarity transformation (\ref{su2v0}), is given by
\be
\vv =  -\fr{1}{2 \sin \theta}\left( \begin{array}{c}
     -\left[a(r)\,e^{rs} + b(r)\,e^{-rs}\,e^{-i\phi}\right]\sin \theta \\
    a(r)\,e^{rs} \left(\cos\theta -1\right)e^{i\phi} + b(r)\,e^{-rs} \left(\cos\theta+1 \right)
\end{array} \right) .\label{generic}
\ee

Next, chose two vectors $\vv_i$ for $i=1,2$ to  be of the form (\ref{generic}) with constants $a_i$ and $b_i$, such that  the set $\{\vv_1,\vv_2\} $ defines an orthonormal system.
One may readily extract that
\bea
< \vv_i, \vv_j > &=&\fr{1}{2 \sin^2 \theta} \int_{-1}^1\left[ \bar{a}_i \,a_j  \left(1-\cos\theta \right)  e^{2rs} + \bar{b}_i \,b_j
\left(\cos\theta +1\right)  e^{-2rs}  \right ] ds\nonumber\acc
&=&\fr{\sinh(2r)}{2r} \left[ \fr{\bar{a}_i \,a_j}{\left( 1+u\right)} +\fr{\bar{b}_i \,b_j}{
\left(1-u\right)}   \right ]
\eea
using the variables (\ref{polar}).
Then, from the orthonormality conditions (\ref{ortho}) the form of the constants $a_i$ and $b_i$ which are now functions of the variables $(r,u,\psi)$  (up to a gauge choice) can be obtained.
In fact, (\ref{generic})   becomes  (\ref{gen1}) by setting
\bea
a_1=\sqrt{\fr{r}{\sinh 2r}}\,\psi, \hs\hs\ &&  b_1=-\sqrt{\fr{r}{\sinh 2r}}\,\sqrt{|\psi|^2},\nonumber\\
 a_2=\sqrt{\fr{r}{\sinh 2r}}\left(1+u\right),\ && b_2=\sqrt{\fr{r}{\sinh 2r}}\,\sqrt{\fr{\bar{\psi}}{\psi}}\,\left(1-u\right).
 \eea
Finally, the associated Higgs field and gauge potentials are given by (\ref{hig1}).
For example, the elements of the Higgs field  are of the form
\bea
\Phi_{ij}&= &-\fr{i}{2 } \int_{-1}^1 s \left[\fr{ \bar{a}_i\, a_j}{  \left(1+u\right)} \, e^{2rs} +
\fr{\bar{b}_i \,b_j}{
\left(1-u\right)} \, e^{-2rs}  \right ] ds\nonumber\acc
&=&-\fr{i}{2}\left(\fr{-\sinh 2r +2r\cosh 2r}{2r^2}\right)\left[\fr{ \bar{a}_i\, a_j}{  \left(1+u\right)}  -
\fr{\bar{b}_i \,b_j}{
\left(1-u\right)}   \right ].
\eea

\subsection{$SU(3)$ Case}

The next non-trivial case, which can be  easily treated is the $SU(3)$ gauge.
Recall that  the solution of the diagonal case has the form
\be
\vv_0 = \left( \begin{array}{c}
      h_1(r,s)\\
      h_2(r,s)
\end{array} \right) \otimes  \left( \begin{array}{c}
      g_1(r,s)\\
      g_2(r,s)
\end{array} \right) = \left( \begin{array}{c}
      v_1(r,s)\\
      w_1(r,s)\\
      v_2(r,s)\\
      w_2(r,s)
\end{array} \right),
\ee
where $v_i=g_1\, h_i $ and $w_i=g_2\, h_i $ for $i=1,2$.
The entries $u_i$ and $w_i$ are known explicitly due to  \cite{doikou-ioannidou}:
\bea
v_1 = a(r)\,\sqrt{s} \,e^{rs},\hs\hs\hs \hs\hs\ \ \,& &\hs w_1 = {W \over s},\nonumber\\
v_2 = \dot W + \left(r -{1\over 2s}\right) W,\hs\hs\hs &&\hs w_2= b(r)\,\sqrt{s}\, e^{-rs}\label{uw}
\eea
where
\bea
W = c(r) \,M\left({1\over 2}, 1; 2rs\right)= c(r) {\left(\sinh(rs) - rs \,e^{-rs} \over \sqrt{rs}\right)}.\label{W}
\eea
$M(k,m;z)$ is the  Whittaker function of first type (see also \cite{doikou-ioannidou}) and  $\dot W$ denotes  the derivative of the function $W$  with respect to $s$.

After performing the inverse similarity transformation the solution of the full Weyl equation becomes
{\small \be
\vv = \fr{1}{4 \sin^2 \theta}\left( \begin{array}{c}
      -\left(h_1 +h_2\, e^{-i \phi}\right)\sin \theta\\
    h_1  \left(\cos \theta -1\right) \,e^{i\phi }  + h_2 \left(\cos \theta +1\right)
\end{array} \right) \otimes
  \left( \begin{array}{c}
      -\left(g_1 + g_2\,e^{-i \phi}\right)\sin \theta\\
      g_1\left(\cos \theta -1\right) e^{i\phi } +g_2\left(\cos \theta +1\right)
\end{array}
\right).\label{f}
\ee}

As before, we choose three distinct vectors $\vv_j$ for $j=1,2,3$ of the form (\ref{f}).
Then, the  associated solutions $v_i^{(j)}, w_i^{(j)}$ are given in terms  of the  constants $a_j, b_j$ and  $c_j$  by  (\ref{uw}) and (\ref{W}).
Again, one may readily extract that  their norm is equal to
 \bea
<\vv_i, \vv_j>& \!\!\! =  \!\!\! &\fr{1}{4\sin^2 \theta}\!\int_{0}^3 \left[{\left(\cos \theta -1\right)^2 \over \sin^2 \theta} v_1^{*(i)}v_1^{(j)} \!+\!
{\left(\cos \theta + 1\right)^2 \over \sin^2 \theta} w_2^{*(i)} w_2^{(j)}\! +\!  v_2^{*(i)}v_2^{(j)} +
w_1^{*(i)}w_1^{(j)}\right ]ds\nonumber\acc
&  \!\!\!\!\!\!=  \!\!\!\!\! & {1\over 16}\left[\bar{a}_i \,a_j\,\fr{q_0}{r^2\, \left(1+u\right)^2}
+\bar{b}_i\, b_j\, \fr{q_1}{r^2\,\left(1-u\right)^2} +\bar{c}_i\,c_j \, \fr{q_0\,q_1}{9r\,|\psi|^2}\right],
\eea
where  $q_0= 1+\left(-1+6r\right)e^{-6r}$, $q_1=1-\left(1+6r\right)e^{6r}$ and  $(r,u,\psi)$ are the parameters defined in  (\ref{polar}).
Note that, for convenience the limits of integration have been shifted in this case to $0$ and $3$.

Requiring the orthonormality conditions (\ref{ortho}) one may fix, not uniquely,
the constants which are now functions of $(r,u,\psi)$ and determine the vectors $\vv_j$.
It is straightforward to show that the constants $a_j$, $b_j$, and $c_j$ are of the form:
\bea
&&\!\!\!\!\!\!\!\!  a_1 = {4\,r \over \sqrt{3\,q_0}} \left(u+1\right) \,e^{i \alpha},\hs \hs\ \ \,
 b_1 ={4r \over \sqrt{3\,q_1}} \left(1-u\right)\, e^{i \beta}, \hs\hs\,\,
  c_1 = {4\,  \sqrt{r} \over \sqrt{3\,q_0\,q_1}}\,\psi\, e^{i\gamma},\hs\hs\nonumber\\
 &&\!\!\!\! \!\!\!\! a_2 = {4\,r \over \sqrt{3\,q_0}} \left(u+1\right) \,e^{i \alpha -{2\pi i \over 3}}, \hs\
 b_2 ={4\,r \over \sqrt{3\,q_1}} \left(1-u\right)\,e^{i \beta+ {2\pi i \over 3}}, \ \ \,\,
 c_2 = {4\,  \sqrt{r} \over \sqrt{3\,q_0\,q_1}}\,  \psi \, e^{i\gamma},\hs\hs\nonumber\\
&&\!\!\!\! \!\!\!\! a_3 = {4\,r \over \sqrt{3\,q_0}}  \left(u+1\right) \, e^{i \alpha-{\pi i \over 3}}, \hs\ \
 b_3 ={4\,r \over \sqrt{3\,q_1}} \left(1-u\right)\, e^{i \beta+{\pi i \over 3}}, \hs\
 c_3 = {4\, \sqrt{r} \over \sqrt{3\,q_0\,q_1}}\, \psi \,e^{i\gamma - \pi i}.\hs\hs
\eea
where $\alpha$, $\beta$, and $\gamma$ are real free parameters, which for simplicity may be set equal to zero.

Then, the Higgs field and gauge potentials can be recovered from (\ref{Higgs}).
In particular, the elements of the  Higgs field  are equal to
{\small \bea
\Phi_{ij}\!\!\!\!& =& \!\!\!\!-\fr{i}{4 |\psi|^2}\!\int_{0}^3 (s-2)\left[{\left(u -1\right)^2 \over |\psi|^2} \,v_1^{*(i)}v_1^{(j)} +
{\left(u+ 1\right)^2 \over |\psi|^2}\, w_2^{*(i)} w_2^{(j)}+  v_2^{*(i)}v_2^{(j)} +
w_1^{*(i)}w_1^{(j)}\right ]ds\nonumber\acc
\!\!\!\!&=&\!\!\!\!-\fr{i}{4}\left[\fr{\bar{a}_i\,a_j}{(1+u)^2}\, \left(\fr{-1-2r+\left(1-4r+6r^2\right)e^{6r}}{4r^3}\right)
-\fr{\bar{b}_i\,b_j}{(1-u)^2}\left( \fr{-1+2r+\left(1+4r+6r^2\right)e^{-6r}}{4r^3}\right)\right.\nonumber\\
&&\left.\hs -\fr{\bar{c}_i\,c_j}{(1-u^2)}\left(\fr{-36r^2-4+\left(2-3r\right)e^{6r}+\left(2+3r\right)e^{-6r}}{18r}\right)\right].
\eea}

\section{Conclusions}
$SU(n+1)$ spherically symmetric monopole solutions of  the full Weyl equation in the case of minimal symmetry breaking  were obtained.
This was done by implementing azimuthal dependence to the  solutions found in \cite{doikou-ioannidou} via a suitable similarity transformation.
More precisely, the existence of a spin chain-like ($XXX$ chain)  part
with a bulk spin-spin interaction and a boundary term within the Weyl equation was exploited.
In this context,  a similarity transformation exists that turns the diagonal boundary terms to generic non-diagonal ones but, leaves the bulk interaction invariant \cite{annecy-group}.
This is precisely the transformation one utilizes to implement the azimuthal dependence to the spherical symmetric solutions.

In this paper, a particular case that involves the simplest Nahm data is considered
and thus,  the isotropic Heisenberg model ($XXX$ chain) is involved.
For generic Nahm data the situation becomes more complicated, and the anisotropic Heisenerg model ($XXZ$ or $XYZ$ chain) is involved.
In the general situation, such a {\it boundary similarity transformation} that leaves the bulk part unaffected does not exist; thus, more sophisticated methods need to be employed (see, for example, References \cite{chin, doikou}).
This is an intriguing issue, that we hope to address in future investigations.

\appendix
\section{Appendix}

We give two explicit examples of the transformation ${\cal U}$ given by (\ref{UU}) for the $SU(4)$ and $SU(5)$ case.

{\it $SU(4)$ Case}: Here $n=3$ and our formulation leads to
\be
{\cal U}^{-1}=\left(\begin{array}{ccc}
1&-\sqrt{2}\bar{\xi}&\bar{\xi}^2\\
\sqrt{2}\xi& -\left(|\xi|^2-1\right)&-\sqrt{2} \bar{\xi}\\
\xi^2&\sqrt{2}\xi&1
\end{array}\right).
\ee

{\it $SU(5)$ Case}: Here $n=4$ and our formulation leads to
\be
{\cal U}^{-1}=\left(\begin{array}{cccc}
1&-\sqrt{3}\bar{\xi}&\sqrt{3}\bar{\xi}^2&-\bar{\xi}^3\\
\sqrt{2}\xi& -\left(2|\xi|^2-1\right)&\bar{\xi}\left(|\xi|^2-2\right)&\sqrt{3}\bar{\xi}^2\\
\sqrt{3}\xi^2&-\xi\left(|\xi|^2-2\right)&-\left(2|\xi|^2-1\right)&-\sqrt{3}\bar{\xi}\\
\xi^3&\sqrt{3}\xi^2&\sqrt{3}\xi&1
\end{array}\right).
\ee

\end{document}